\documentclass[aps,prb,reprint,superscriptaddress]{revtex4-2}
\usepackage{amsmath, amssymb}
\usepackage{graphicx}
\usepackage{braket}
\begin{document}

\title{A coarse-grained description of anharmonic lattice environments affecting the quantum dynamics of charge carriers}

\author{Kuniyuki Miwa}
    \email{kuniyukimiwa@ims.ac.jp}
    \affiliation{Institute for Molecular Science, National Institutes of Natural Sciences, Okazaki 444-8585, Japan}
    \affiliation{Graduate Institute for Advanced Studies, SOKENDAI, Okazaki 444-8585, Japan}
\author{Souichi Sakamoto}
    \affiliation{Institute for Molecular Science, National Institutes of Natural Sciences, Okazaki 444-8585, Japan}
\author{Ken Funo}
    \affiliation{Institute for Molecular Science, National Institutes of Natural Sciences, Okazaki 444-8585, Japan}
\author{Akihito Ishizaki}
    \email{ishizaki@ims.ac.jp}
    \affiliation{Institute for Molecular Science, National Institutes of Natural Sciences, Okazaki 444-8585, Japan}
    \affiliation{Graduate Institute for Advanced Studies, SOKENDAI, Okazaki 444-8585, Japan}

\date{\today}

\begin{abstract}
Lattice softness has a significant impact on charge carrier dynamics in condensed matter systems, contributing to the emergence of various properties and functions. Examples include the remarkable carrier lifetimes and defect tolerances of hybrid organic--inorganic perovskites. Recent studies suggest the contribution of quartic anharmonicity of the lattice vibrations. The quartic anharmonicity can be discussed with a double-well potential, and the transition between the two minima can be coarse-grained as a two-state jump stochastic process. Such a stochastic approach is typically employed to describe dynamic fluctuations introduced into a system by two-state transitions in the surroundings. To investigate charge transport in materials, however, it is crucial to describe not only the fluctuations but also the dynamic lattice distortion associated with charge transport. Therefore, there is a need for a theory to describe the charge carrier dynamics proceeding alongside the lattice distortion dynamics. In this study, we present a theory that describes quantum dynamics under the influence of an environment with two stable states, termed a bistable environment. The theory describes the effects of fluctuations and dissipation induced from the bistable environment in a reasonable manner, and the effects exhibit a different temperature dependence than the widely employed Gaussian environment. The physical implication of this temperature dependence is provided in terms of the environmental dynamics. The results of this study are expected to provide a step forward in describing charge carrier dynamics in materials with lattice softness and pronounced lattice anharmonicity, \textit{e.g.}, hybrid organic--inorganic perovskites. Moreover, these findings represent an advancement in our understanding of and capacity to predict and control the physical properties and functions of these materials.
\end{abstract}

\maketitle

\section{\label{sec:Intro}Introduction}

Effects of lattice softness on charge carrier dynamics play essential roles in the emergence of various properties and functions of condensed matter systems~\cite{Hendry2004, Gaal2007, Franchini2021}.
Hybrid organic--inorganic perovskites are typical examples of functional materials that have been lately investigated from this perspective~\cite{Zhu2015, Miyata2017a}.
These materials have attracted attention as high-performance solar cell materials with remarkable carrier lifetimes and defect tolerances~\cite{Lee2012, Stranks2013, Shi2015, Chen2016, Herz2017, Jena2019}.
Recent studies suggest the possibility that anharmonicity of lattice vibrations makes a significant contribution~\cite{Beecher2016, Yaffe2017, Miyata2018, Katan2018}.
Extensive studies have been conducted on the relation between the anharmonicity and remarkable properties of the perovskite materials~\cite{Mayers2018, Li2019h, Ferreira2020, Wang2021e, Lanigan-Atkins2021, Tadano2022, Schilcher2023a, Fransson2023}.
However, no sufficient understanding has been obtained on the quantum dynamics of charge carriers under the influence of non-Gaussian fluctuations originating from anharmonic lattice vibrations.

As a typical example of the anharmonic vibrations, we consider a double-well potential with quartic anharmonicity. Indeed, the contribution of quartically anharmonic vibrations in hybrid organic--inorganic perovskites has been proposed~\cite{Beecher2016, Yaffe2017, Miyata2018, Katan2018}.
The transitions between the two potential minima are well-described as the two-state jump (TSJ) stochastic process~\cite{Kubo1969}.
This is a simple stochastic model for describing the impact of dynamic modulations induced by an environment on the system under study and has been successfully employed to investigate the influences of two-state transitions in the surrounding environment, such as chemical exchange~\cite{Migchelsen1973, Jeener1979, Sanda2006, Jansen2007} and spectral broadening~\cite{Kubo1969}.
On the other hand, alongside the dynamic fluctuations, the dynamics of lattice distortion associated with charge transport are crucial for comprehending charge transport in materials~\cite{Tamura2008, Tamura2013, Smith2014, Smith2015, Kato2018b, Yan2018, Yan2019, Fetherolf2020, Balzer2022, Fetherolf2023}.
Polaron formation is a typical example, and lattice distortion significantly affects charge mobility. This highlights the need to develop a theory that describes charge dynamics as it proceeds along with lattice distortion dynamics, because the conventional stochastic approaches typically disregard the back-actions of the system on the environment, such as energy dissipation.

To tackle this issue, we address the Nakajima--Zwanzig equation~\cite{Nakajima1958, Zwanzig1960} with the second-order perturbative truncation in terms of system--environment coupling. This is also termed the second-order perturbative time-convolution quantum master equation (TC2-QME). This equation can be recast into a set of simultaneous equations, indicating that an environment induces the TSJ noise in the system regardless of what statistical properties we assume for the environment~\cite{Ishizaki2010}.
The connection between the TC2-QME and TSJ noise is also reported in Ref.~\onlinecite{Yoon1975}. Moreover, this equation includes dissipative terms; therefore, it can be expected to adequately describe environmental dynamics, \textit{i.e.}, lattice distortion dynamics, in response to the quantum systems such as charge carriers.

In this study, we present a theory to describe the dynamics of a quantum system coupled to an environment with two stable states, termed a bistable environment. We show that the theory describes the effects of fluctuations and dissipation induced from the bistable environment in a reasonable manner. The fluctuations and dissipation exhibit a temperature dependence different from that in the case of the widely employed Gaussian environment. The physical interpretation of this temperature dependence and difference between the bistable and Gaussian environments are provided in terms of the environmental reorganization process.

\section{\label{sec:Model}Model}

In this study, we identify the electronic degrees of freedom (DOFs) as the relevant system and treat the lattice DOFs as the environment. To describe charge carriers coupled to the environment consisting of anharmonic lattice vibrations, we decompose the total Hamiltonian as
$\hat{H} = \hat{H}_{\rm el} + \hat{H}_{\rm env} + \hat{H}_{\rm el-env}$,
where $\hat{H}_{\rm el}$ and $\hat{H}_{\rm env}$ describe the electronic and environmental DOFs, respectively. The couplings between the electronic and environmental DOFs are expressed by $\hat{H}_{\rm el-env}$. Explicit expressions for the individual Hamiltonians are provided below.

We model charge carriers to construct the electronic Hamiltonian, $\hat{H}_{\rm el}$ by employing the tight-binding model with one state per site. The number of sites is assumed to be $N$, and hence the electronic Hamiltonian is given as
\begin{equation}
    \hat{H}_{\rm el}
    =
    \sum_{n=1}^N \varepsilon_n \hat{c}^\dagger_n \hat{c}_n
    -
    \sum_{n \neq n'}^N v_{nn'}
    (\hat{c}^\dagger_n \hat{c}_{n'} + \hat{c}^\dagger_{n'} \hat{c}_n).
    \label{eq:H_el}
\end{equation}
In this equation, $\hat{c}^\dagger_n$ and $\hat{c}_n$ are the creation and annihilation operators of a charge carrier at site $n$ with energy $\varepsilon_n$, respectively. The hopping integral between sites $n$ and $n'$ is denoted by $v_{nn'}$. We focus on the manifold of a single charge carrier and consider the Hilbert space spanned by the basis set of $\ket{n} = \hat{c}_n^\dagger \ket{\rm G}$, each of which represents a charge created on site $n$ against the charge vacuum state, $\ket{\rm G}$, \textit{e.g.} the electronic ground state for photogenerated charge carriers.

\begin{figure}
\includegraphics[bb=0 0 244 85]{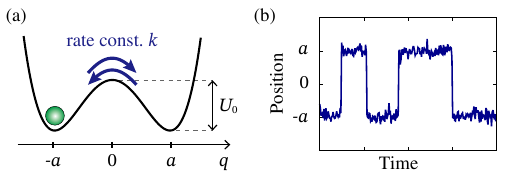}
\caption{\label{fig:SchematicIllustration}
(a) Schematic illustration of an anharmonic oscillator in the symmetric double-well potential defined by Eq.~\eqref{eq:potential}.
(b) Time trajectory of the position of the oscillator.
}
\end{figure}

The environmental DOFs are explored to construct the environmental Hamiltonian, $\hat{H}_{\rm env}$. The presence of anharmonic modes and couplings to charge carriers have been reported in the literature~\cite{Yang2017b, Marronnier2017, Yang2020c, Wang2022b, Zacharias2023a}.
These modes show bistable characteristics in the potential energy surface. For simplicity, we treat the lattice vibrations as independent anharmonic modes. Note that the neglect of coupling among the anharmonic modes cannot be justified in general. However, a similar assumption was made in elucidating the dielectric properties of barium titanate~\cite{Slater1950}.
We thus adopt this assumption as an initial step. We let $q_m$ be the displacement of the $m$-th mode from its equilibrium position. Hence, the environmental Hamiltonian is given as an ensemble of anharmonic oscillators, each of which is characterized by a symmetric double-well potential:
\begin{align}
    \phi(q) 
    = 
    U_0 \left( 1 - \frac{q^2}{a^2} \right)^2,
    \label{eq:potential}
\end{align}
where the two minima are located at $q = \pm a$ and separated by the barrier $U_0$ at $q = 0$, as shown in Fig.~\ref{fig:SchematicIllustration}(a). Figure~\ref{fig:SchematicIllustration}(b) shows that the trajectory can be coarse-grained and viewed as a TSJ process. We therefore describe the transition between the two potential minima as the TSJ stochastic process~\cite{Kubo1969} with a temperature-dependent rate constant:
\begin{align}
    k_{\rm TSJ}
    \propto 
    \exp(-\beta U_0)
    \label{eq:jump-rate}
\end{align}
with $\beta = 1 / k_{\rm B} T$, where $k_{\rm B}$ and $T$ are the Boltzmann constant and temperature, respectively. The classical correlation function of the coordinates is thus approximated as
\begin{align}
    S_{m}(t)
    =
    \langle q_{m}(t) q_{m}(0) \rangle_{\rm env}
    \simeq
    a^2 
    \exp(-2 k_{\rm TSJ} t).
    \label{eq:correlation-func}
\end{align}
It should be noted that the amplitude of this correlation function is independent of temperature, in contrast to that for the thermal fluctuations.

To model the electron--environment coupling, we consider the electrostatic potential between a charge carrier occupying the $m$-th mode and the $n$-th site. For simplicity, we employ the following assumptions: the charge carrier is represented by the point charge found on site $n$; and the effects of screening by other charge carriers and ions are neglected. We expand on-site energy with respect to the displacements $\{q_m\}$ up to the first order, leading to
\begin{align}
    \hat{H}_{\rm el-env}
    &=
    - \sum_{nm} f_{nm} q_{m} \hat{c}_n^\dagger \hat{c}_n
    \label{eq:H_el-env}
\end{align}
with $f_{nm}$ being the coupling strength. Equations~\eqref{eq:H_el}, \eqref{eq:correlation-func}, and \eqref{eq:H_el-env} suggest that the modulated electronic energy takes discrete values, $\varepsilon_n - \sum_m f_{nm} q_m$ with $q_m = \pm a$, and that the modulation amplitude is independent of temperature. We note that by extending our model, the effects of fluctuations in the hopping integral between electronic sites can be investigated in a manner similar to that used in the present study.

\section{\label{sec:EOM}Equations of Motion}

An adequate description of the quantum dynamics of the charge carrier is given by the reduced density operator, that is, the partial trace of the total density operator over the environmental DOFs, $\hat\rho(t) = {\rm Tr}_{\rm env} \hat\rho_{\rm tot}(t)$. To obtain an equation of motion for the reduced density operator under the influence of the bistable environment, we address the TC2-QME~\cite{Nakajima1958, Zwanzig1960}.
As demonstrated in Ref.~\onlinecite{Ishizaki2010}, the TC2-QME with the Drude--Lorentz spectral density can be recast into a simultaneous equation consisting of two equations per site. This structure corresponds to the dynamics of a quantum system modulated by TSJ stochastic noise~\cite{Kubo1969}.
Indeed, the spectral lineshapes computed with the TC2-QME exhibit peak splitting even if the Gaussian environment is assumed~\cite{Ishizaki2010}.
This splitting indicates that the environment in the TC2-QME possesses only two states. It should be noted that the TC2-QME includes dissipative terms, whereas the TSJ model is generally employed as the classical stochastic process in the absence of dissipation~\cite{Kubo1969}.
Therefore, the TC2-QME is anticipated to provide a relevant description of the dynamics of a quantum system coupled to the bistable environment.

The relaxation kernel of the TC2-QME is composed of the symmetrized correlation function and the response function of the environmental coordinates, Eqs.~(4.16) and (4.17) of Ref.~\onlinecite{Ishizaki2010}. In the TSJ model, the amplitude of the classical two-body correlation function of $q_m$ is temperature independent, as shown in Eq.~\eqref{eq:correlation-func}. Therefore, we assume that the symmetrized correlation function is identical to the classical two-body correlation function, Eq.~\eqref{eq:correlation-func}. Regarding the response function, its Fourier--Laplace transform, which is also termed the spectral density, can be obtained through the quantum fluctuation--dissipation relation (FDR) for environmental DOFs~\cite{Kubo1991},
\begin{equation}
    {\rm Im} \chi_m[\omega]
    =
    \frac{2\pi}{\hbar} \tanh\frac{\beta\hbar\omega}{2} S_m[\omega],
    \label{eq:FDR1}
\end{equation}
where $S_{m} [\omega]$ is the Fourier transform of the symmetrized correlation function, $S_m[\omega] = 4 a^2 k_{\rm TSJ} / ( \omega^2 + 4 k_{\rm TSJ}^2 )$. It should be noted that the spectral density can be temperature-dependent in the case of non-Gaussian environments such as a spin bath, as discussed in Ref.~\onlinecite{Weiss2012}. Consequently, the response function is obtained as
\begin{align}
    \chi_{m}(t)
    =
    \frac{4}{\hbar} \int^\infty_0 d\omega
    \tanh\frac{\beta \hbar \omega}{2} S_{m}[\omega] \sin\omega t.
    \label{eq:response-func}
\end{align}
In the high-temperature limit characterized by $\beta\hbar\omega \ll 1$, the response function is approximated as
$\chi_m(t) \simeq 2 a^2 k_{\rm TSJ} (k_{\rm B} T )^{-1} e^{-2k_{\rm TSJ}t}$,
indicating that the dissipative influence decreases with increasing temperature. The physical implication of this anomalous temperature dependence will be discussed in connection with Fig.~\ref{fig:TRPE}.

By employing Eqs.~\eqref{eq:correlation-func} and \eqref{eq:response-func} for the electron--environment interaction in Eq.~\eqref{eq:H_el-env}, the equation of motion for the reduced density operator is obtained as
\begin{subequations}
\label{eq:EOM}
\begin{align}
    \frac{d}{dt} \hat\rho(t)
    =
    -i \hat{\cal L}_{\rm el} \hat\rho(t)
    +
    \sum_{nn'} i \hat{V}_n^\times \Delta_{nn'}^2
	\left[ \hat\sigma_{n'}(t) - \sum_{j=1}^\infty \hat\mu_{n'j}(t) \right],
\end{align}
where
$\hat{\cal L}_{\rm el} \hat{ \rho } = [ \hat{H}_{\rm el} , \hat{\rho} ]/\hbar$,
$\hat{V}_n = \lvert n \rangle\langle n \rvert$,
$\hbar^2\Delta_{nn'}^2 = \sum_m f_{nm} f_{n' m} a^2$,
and the super-operator notations,
$\hat{O}^{\times} \hat{f} = \hat{O} \hat{f} - \hat{f} \hat{O}$
and
$\hat{O}^{\circ} \hat{f} = \hat{O} \hat{f} + \hat{f} \hat{O}$,
have been introduced. The auxiliary operators, $\hat{ \sigma}_n(t)$ and $\hat{\mu}_{nj}(t)$ describe the environmental TSJ process and quantum corrections to the high-temperature limit of the response function, respectively. Their time evolution is given by
\begin{gather}
    \frac{d}{dt} \hat\sigma_n(t)
    =
    -( i \hat{\cal L}_{\rm el} + 2 k_{\rm TSJ} ) \hat\sigma_n(t)
    +
    i ( \hat{V}_n^\times - i {\cal Z} \hat{V}_n^\circ)
    \hat\rho(t),
    \\
    \frac{d}{dt} \hat\mu_{nj}(t)
    =
    -\left( i \hat{\cal L}_{\rm el} + \nu_j - \frac{\nu_1}{2} \right) \hat\mu_{nj}(t)
    +
    \zeta_j \hat{V}_n^\circ \hat\rho(t),
\end{gather}
\end{subequations}
where $\nu_j$ is the $j$-th Matsubara frequency written as $\nu_j = 2 \pi j / \beta \hbar$, and we have introduced
$\zeta_j = 8 k_{\rm TSJ} / \{ [ (\nu_j - \nu_1/2)^2 - 4 k_{\rm TSJ}^2 ] \beta \hbar \}$
and
${\cal Z}=\sum_{j=1}^\infty \zeta_j$.

\section{\label{sec:Results}Results and Discussion}

\begin{figure}
\includegraphics[bb=0 0 244 142]{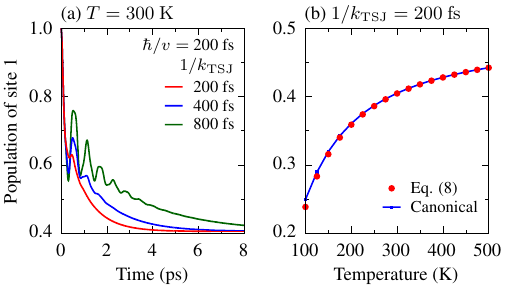}
\caption{
\label{fig:population}
(a) Time evolution of the population of site 1 in a two-site system calculated with Eq.~\eqref{eq:EOM} for several values of the jump rate, $k_{\rm TSJ}$. In this panel, the temperature is set as $T = 300\,{\rm K}$.
(b) Comparison between the converged values of the time-evolved population (red circles) and the canonical distribution (blue squares) of the population of site 1 as a function of $T$. The energy difference between sites 1 and 2, $\varepsilon_1 - \varepsilon_2$ and the electron hopping integral between sites, $v$ are set to $\varepsilon_1 - \varepsilon_2 = 10\,{\rm meV}$ and $v = 3.29\,{\rm meV}~(\hbar / v = 200\,{\rm fs})$, respectively; the electron--environment coupling strengths, $\hbar \Delta_{nn'} = (\sum_m f_{nm} f_{n'm} a^2)^{1/2}$ are set to $\hbar \Delta_{11} = \hbar \Delta_{22} = 5\,{\rm meV}$ and $\hbar \Delta_{12} = \hbar \Delta_{21} = 0$. In panel (b), the jump rate is set to $k_{\rm TSJ}^{-1} = 200\,{\rm fs}$.
}
\end{figure}

To show that the theory is capable of adequately describing quantum dynamics in the bistable environment, we investigate the dynamics of a photogenerated charge carrier. Because the TC2-QME satisfies the FDR for environmental DOFs, the present theory is expected to produce the thermal equilibrium at finite temperature. To confirm this, we consider the population dynamics in a two-site system for demonstration purposes. Figure~\ref{fig:population}(a) shows the time evolution of the population of site 1 calculated with Eq.~\eqref{eq:EOM} for several values of jump rate, $k_{\rm TSJ}$. As the initial condition, only site 1 is excited. The site 1 population converges to a steady value over time. Figure~\ref{fig:population}(b) displays the converged values and canonical distribution of the site 1 population as a function of temperature. Both are in good agreement with each other in the temperature range shown in the figure.


\begin{figure}
\includegraphics[bb=0 0 244 142]{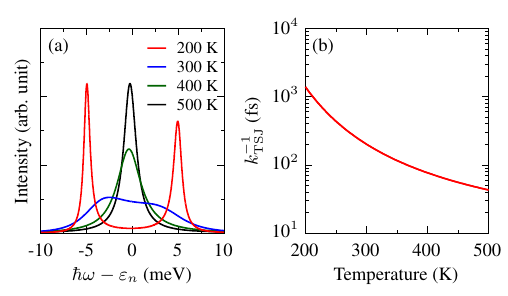}
\caption{
\label{fig:spectrum}
(a) Spectral function of a single-site system calculated with Eq.~\eqref{eq:EOM} for various values of temperature $T$. The electron--environment coupling strength is set to $\hbar \Delta_{nn} = 5\,{\rm meV}$ ($\Delta_{nn}^{-1} = 131\,{\rm fs}$). (b) Temperature dependence of the jump rate. The jump rate at $300\,{\rm K}$ and energy barrier are set to $k_{\rm TSJ}^{-1} = 200\,{\rm fs}$ and $U_0 = 100\,{\rm meV}$, respectively.
}
\end{figure}

We show that the theory can adequately describe the fluctuations introduced by the bistable environment. To this end, we analyze the spectral function, $A_n (\omega) = -(1/\pi) {\rm Im} \int_0^\infty dt\, e^{ i \omega t } g_n^{\rm r}(t)$ with $g_n^{\rm r}(t) = \theta(t) {\rm Tr}[ \{ \hat{c}_n(t) , \hat{c}^\dagger_n(0) \} \hat{\rho}_{\rm tot}(0)] / i \hbar$, which provides information on the probability of adding or removing a particle of energy $\hbar \omega$~\cite{Coleman2015}.
The information content of this function is essentially identical to that of the linear absorption lineshape of exciton systems. Because the transition energy is modulated by local fluctuations in the surroundings, the lineshape provides an important window into the structure and dynamics of the environment~\cite{Kubo1969}.
Figure~\ref{fig:spectrum}(a) shows the spectral function of a single-site system at various temperatures. As shown in Eq.~\eqref{eq:jump-rate} and Fig.~\ref{fig:spectrum}(b), the jump rate depends on temperature. At lower temperatures so that the jump rate $k_{\rm TSJ}$ satisfies $\Delta_{nn} / k_{\rm TSJ} \gg 1$, the spectral function is peaked at two resonance energies, $\hbar \omega = \varepsilon_n \pm \hbar \Delta_{nn}$. At higher temperatures where $\Delta_{nn} / k_{\rm TSJ} \ll 1$ is satisfied, on the other hand, the spectral function exhibits a single Lorentzian centered at $\hbar \omega = \varepsilon_n$. This temperature dependence corresponds to the characteristics of the TSJ process in the Anderson--Kubo stochastic theory~\cite{Anderson1954, Kubo1954}.


\begin{figure}
\includegraphics[bb=0 0 244 142]{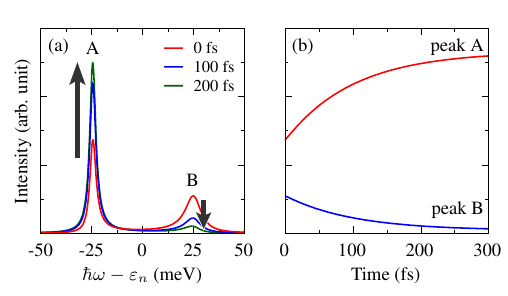}
\caption{\label{fig:TRPE}
(a) Time-resolved emission spectrum of a single-site system calculated with Eq.~\eqref{eq:EOM} for several values of the delay time $t$. (b) Time evolution of the high- and low-energy peak intensities. The parameters are set to $\hbar \Delta_{nn} = 25\,{\rm meV}$, $k_{\rm TSJ}^{-1} = 200\,{\rm fs}$, and $T = 300\,{\rm K}$.
}
\end{figure}

To demonstrate the capability of the theory to describe the back-action of the system on the environment, we display the time-resolved emission spectra of a single-site system for various values of the delay time after photoexcitation in Fig.~\ref{fig:TRPE}~\cite{Ishizaki2009a}.
For the spectra to show two peaks, we consider a case that satisfies $\Delta_{nn} / k_{\rm TSJ} \gg 1$. The intensity of the low-energy peak increases, whereas that of the high-energy peak decreases, with time $t$. The time constant of the evolution is estimated to be $0.01\,{\rm fs}^{-1}$, which is twice the jump rate, $2k_{\rm TSJ}$.


\begin{figure}
\centering
\includegraphics[bb=0 0 230 265]{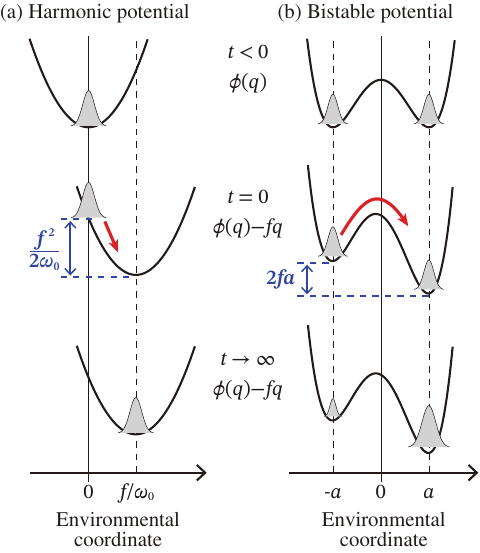}
\caption{\label{fig:Reorganication}
Schematic illustration of the environmental reorganization in the cases of the (a) harmonic potential, $\phi (q) = (1/2)\omega_0 q^2$ and (b) bistable potential, $\phi (q) = U_0 ( 1 - q^2 / a^2 )^2$. The packets depict the distributions of the environmental states, and the red arrows illustrates the reorganization dynamics. A field, $-fq$, is added to the potential at time $t=0$. In the harmonic case, the field causes only a horizontal shift of the potential, and the reorganization takes place with dissipation of reorganization energy, $E_{\rm R} = f^2/2\omega_0$, independent of temperature. In the bistable case, the field causes the energy difference between the left and right wells, $\Delta E = 2fa$; however, the distributions in the left and right wells at $t=0$ are equal. Hence, the reorganization proceeds to the equilibrium with the canonical distributions. 
}
\end{figure}

The time evolution of the spectrum shown in Fig.~\ref{fig:TRPE} can be understood in terms of the environmental reorganization, in which the environmental DOFs relax from a non-equilibrium state to the equilibrium state by dissipating the so-called reorganization energy. This process corresponds to the relaxation function~\cite{Kubo1991, Ishizaki2010}
\begin{align}
    \Gamma_{m}(t) 
    = 
    \int^\infty_t ds \,\chi_{m}(s) 
    \simeq 
    \frac{a^2}{k_{\rm B} T} 
    \exp(-2k_{\rm TSJ}t).
    \label{eq:relaxation-func}
\end{align}
A charge carrier is generated by the photoexcitation of the electronic system at time $t=0$, and a field that linearly depends on the environmental coordinate is added to the potential, \textit{i.e.}, $\phi(q) - \theta(t) fq$ with $\theta(t)$ being the Heaviside step function. In the case of the harmonic potential with frequency $\omega_0$ shown in Fig.~\ref{fig:Reorganication}(a), the field causes only a horizontal shift of the potential, and the reorganization takes place with the dissipation of reorganization energy, $E_{\rm R} = f^2/2\omega_0$. This causes a shift of the maximum peak position of the emission spectrum to the lower-energy side as time progresses. In the bistable case in Fig.~\ref{fig:Reorganication}(b), on the other hand, the field causes the energy difference between the left and right wells, $\Delta E = 2fa$; however, the distributions in the left and right wells at $t=0$ are equal, $P_{\rm L}(t=0) = P_{\rm R}(t=0) = 1/2$. Hence, the reorganization proceeds to the equilibrium with the canonical distributions $P_{\rm L}(t\to\infty) = e^{-\beta \Delta E}/(1+e^{-\beta \Delta E})$ and $P_{\rm R}(t\to \infty) = 1/(1+e^{-\beta \Delta E})$. This leads to changes in peak intensities but not in peak positions of the spectrum because the environmental coordinate is coarse-grained to take only two discrete values, $q = \pm a$. Furthermore, it is noteworthy that the degree of the redistribution, $\Delta P_{\rm L/R} = P_{\rm L/R}(t=0)- P_{\rm L/R}(t\to \infty)$, decreases with increasing temperature. As a result, the environmental reorganization is less likely to occur at higher temperatures, yielding less dissipative influence with increasing temperature. This provides the physical implication of the temperature dependence of the response function given by Eq.~\eqref{eq:response-func}.

\section{\label{sec:Conclusion}Conclusion}

We presented a theory to describe the dissipative dynamics of a quantum system coupled to a bistable environment. As demonstrated with the numerical results, the theory can describe the effects of both the fluctuations and dissipation introduced by the bistable environment in a consistent manner. The fluctuations and dissipation exhibit a different temperature dependence than the widely employed Gaussian environment. In particular, the dissipative influence decreases with increasing temperature. The physical implications of this temperature dependence are analyzed in terms of the environmental reorganization process. The description of the dissipative influence as well as the fluctuations is crucial for understanding the charge carrier dynamics because lattice distortion in response to charge carriers is inherently unavoidable in materials, especially materials with lattice softness. The theory presented in this study is expected to provide a step forward in describing charge carrier dynamics in materials with lattice softness and pronounced lattice anharmonicity, such as hybrid organic--inorganic perovskites; in addition, it contributes to advances in our understanding of and ability to predict and control the physical properties and functions of such materials. Furthermore, the insights gained in this study are expected to contribute to the comprehension of open quantum dynamics in non-Gaussian environments~\cite{Prokofev2000, Hsieh2018b, Hsieh2018c, Funo2023}.

\begin{acknowledgments}
K. M. acknowledges support from JSPS KAKENHI (Grant Number 21K14481).
K. F. acknowledges support from JSPS KAKENHI (Grant Number 23K13036).
This work was also supported by JSPS KAKENHI (Grant Number 21H01052) and the MEXT Quantum Leap Flagship Program (Grant Number JPMXS0120330644).
\end{acknowledgments}


\begin{thebibliography}{58}%
\makeatletter
\providecommand \@ifxundefined [1]{%
 \@ifx{#1\undefined}
}%
\providecommand \@ifnum [1]{%
 \ifnum #1\expandafter \@firstoftwo
 \else \expandafter \@secondoftwo
 \fi
}%
\providecommand \@ifx [1]{%
 \ifx #1\expandafter \@firstoftwo
 \else \expandafter \@secondoftwo
 \fi
}%
\providecommand \natexlab [1]{#1}%
\providecommand \enquote  [1]{``#1''}%
\providecommand \bibnamefont  [1]{#1}%
\providecommand \bibfnamefont [1]{#1}%
\providecommand \citenamefont [1]{#1}%
\providecommand \href@noop [0]{\@secondoftwo}%
\providecommand \href [0]{\begingroup \@sanitize@url \@href}%
\providecommand \@href[1]{\@@startlink{#1}\@@href}%
\providecommand \@@href[1]{\endgroup#1\@@endlink}%
\providecommand \@sanitize@url [0]{\catcode `\\12\catcode `\$12\catcode
  `\&12\catcode `\#12\catcode `\^12\catcode `\_12\catcode `\%12\relax}%
\providecommand \@@startlink[1]{}%
\providecommand \@@endlink[0]{}%
\providecommand \url  [0]{\begingroup\@sanitize@url \@url }%
\providecommand \@url [1]{\endgroup\@href {#1}{\urlprefix }}%
\providecommand \urlprefix  [0]{URL }%
\providecommand \Eprint [0]{\href }%
\providecommand \doibase [0]{https://doi.org/}%
\providecommand \selectlanguage [0]{\@gobble}%
\providecommand \bibinfo  [0]{\@secondoftwo}%
\providecommand \bibfield  [0]{\@secondoftwo}%
\providecommand \translation [1]{[#1]}%
\providecommand \BibitemOpen [0]{}%
\providecommand \bibitemStop [0]{}%
\providecommand \bibitemNoStop [0]{.\EOS\space}%
\providecommand \EOS [0]{\spacefactor3000\relax}%
\providecommand \BibitemShut  [1]{\csname bibitem#1\endcsname}%
\let\auto@bib@innerbib\@empty
\bibitem [{\citenamefont {Hendry}\ \emph {et~al.}(2004)\citenamefont {Hendry},
  \citenamefont {Wang}, \citenamefont {Shan}, \citenamefont {Heinz},\ and\
  \citenamefont {Bonn}}]{Hendry2004}%
  \BibitemOpen
  \bibfield  {author} {\bibinfo {author} {\bibfnamefont {E.}~\bibnamefont
  {Hendry}}, \bibinfo {author} {\bibfnamefont {F.}~\bibnamefont {Wang}},
  \bibinfo {author} {\bibfnamefont {J.}~\bibnamefont {Shan}}, \bibinfo {author}
  {\bibfnamefont {T.~F.}\ \bibnamefont {Heinz}},\ and\ \bibinfo {author}
  {\bibfnamefont {M.}~\bibnamefont {Bonn}},\ }\href
  {https://doi.org/10.1103/PhysRevB.69.081101} {\bibfield  {journal} {\bibinfo
  {journal} {Phys. Rev. B}\ }\textbf {\bibinfo {volume} {69}},\ \bibinfo
  {pages} {081101} (\bibinfo {year} {2004})}\BibitemShut {NoStop}%
\bibitem [{\citenamefont {Gaal}\ \emph {et~al.}(2007)\citenamefont {Gaal},
  \citenamefont {Kuehn}, \citenamefont {Reimann}, \citenamefont {Woerner},
  \citenamefont {Elsaesser},\ and\ \citenamefont {Hey}}]{Gaal2007}%
  \BibitemOpen
  \bibfield  {author} {\bibinfo {author} {\bibfnamefont {P.}~\bibnamefont
  {Gaal}}, \bibinfo {author} {\bibfnamefont {W.}~\bibnamefont {Kuehn}},
  \bibinfo {author} {\bibfnamefont {K.}~\bibnamefont {Reimann}}, \bibinfo
  {author} {\bibfnamefont {M.}~\bibnamefont {Woerner}}, \bibinfo {author}
  {\bibfnamefont {T.}~\bibnamefont {Elsaesser}},\ and\ \bibinfo {author}
  {\bibfnamefont {R.}~\bibnamefont {Hey}},\ }\href
  {https://doi.org/10.1038/nature06399} {\bibfield  {journal} {\bibinfo
  {journal} {Nature}\ }\textbf {\bibinfo {volume} {450}},\ \bibinfo {pages}
  {1210} (\bibinfo {year} {2007})}\BibitemShut {NoStop}%
\bibitem [{\citenamefont {Franchini}\ \emph {et~al.}(2021)\citenamefont
  {Franchini}, \citenamefont {Reticcioli}, \citenamefont {Setvin},\ and\
  \citenamefont {Diebold}}]{Franchini2021}%
  \BibitemOpen
  \bibfield  {author} {\bibinfo {author} {\bibfnamefont {C.}~\bibnamefont
  {Franchini}}, \bibinfo {author} {\bibfnamefont {M.}~\bibnamefont
  {Reticcioli}}, \bibinfo {author} {\bibfnamefont {M.}~\bibnamefont {Setvin}},\
  and\ \bibinfo {author} {\bibfnamefont {U.}~\bibnamefont {Diebold}},\ }\href
  {https://doi.org/10.1038/s41578-021-00289-w} {\bibfield  {journal} {\bibinfo
  {journal} {Nat. Rev. Mater.}\ }\textbf {\bibinfo {volume} {6}},\ \bibinfo
  {pages} {560} (\bibinfo {year} {2021})}\BibitemShut {NoStop}%
\bibitem [{\citenamefont {Zhu}\ and\ \citenamefont {Podzorov}(2015)}]{Zhu2015}%
  \BibitemOpen
  \bibfield  {author} {\bibinfo {author} {\bibfnamefont {X.-Y.}\ \bibnamefont
  {Zhu}}\ and\ \bibinfo {author} {\bibfnamefont {V.}~\bibnamefont {Podzorov}},\
  }\href {https://doi.org/10.1021/acs.jpclett.5b02462} {\bibfield  {journal}
  {\bibinfo  {journal} {J. Phys. Chem. Lett.}\ }\textbf {\bibinfo {volume}
  {6}},\ \bibinfo {pages} {4758} (\bibinfo {year} {2015})}\BibitemShut
  {NoStop}%
\bibitem [{\citenamefont {Miyata}\ \emph {et~al.}(2017)\citenamefont {Miyata},
  \citenamefont {Atallah},\ and\ \citenamefont {Zhu}}]{Miyata2017a}%
  \BibitemOpen
  \bibfield  {author} {\bibinfo {author} {\bibfnamefont {K.}~\bibnamefont
  {Miyata}}, \bibinfo {author} {\bibfnamefont {T.~L.}\ \bibnamefont
  {Atallah}},\ and\ \bibinfo {author} {\bibfnamefont {X.-Y.}\ \bibnamefont
  {Zhu}},\ }\href {https://doi.org/10.1126/sciadv.1701469} {\bibfield
  {journal} {\bibinfo  {journal} {Sci. Adv.}\ }\textbf {\bibinfo {volume}
  {3}},\ \bibinfo {pages} {1} (\bibinfo {year} {2017})}\BibitemShut {NoStop}%
\bibitem [{\citenamefont {Lee}\ \emph {et~al.}(2012)\citenamefont {Lee},
  \citenamefont {Teuscher}, \citenamefont {Miyasaka}, \citenamefont
  {Murakami},\ and\ \citenamefont {Snaith}}]{Lee2012}%
  \BibitemOpen
  \bibfield  {author} {\bibinfo {author} {\bibfnamefont {M.~M.}\ \bibnamefont
  {Lee}}, \bibinfo {author} {\bibfnamefont {J.}~\bibnamefont {Teuscher}},
  \bibinfo {author} {\bibfnamefont {T.}~\bibnamefont {Miyasaka}}, \bibinfo
  {author} {\bibfnamefont {T.~N.}\ \bibnamefont {Murakami}},\ and\ \bibinfo
  {author} {\bibfnamefont {H.~J.}\ \bibnamefont {Snaith}},\ }\href
  {https://doi.org/10.1126/science.1228604} {\bibfield  {journal} {\bibinfo
  {journal} {Science}\ }\textbf {\bibinfo {volume} {338}},\ \bibinfo {pages}
  {643} (\bibinfo {year} {2012})}\BibitemShut {NoStop}%
\bibitem [{\citenamefont {Stranks}\ \emph {et~al.}(2013)\citenamefont
  {Stranks}, \citenamefont {Eperon}, \citenamefont {Grancini}, \citenamefont
  {Menelaou}, \citenamefont {Alcocer}, \citenamefont {Leijtens}, \citenamefont
  {Herz}, \citenamefont {Petrozza},\ and\ \citenamefont
  {Snaith}}]{Stranks2013}%
  \BibitemOpen
  \bibfield  {author} {\bibinfo {author} {\bibfnamefont {S.~D.}\ \bibnamefont
  {Stranks}}, \bibinfo {author} {\bibfnamefont {G.~E.}\ \bibnamefont {Eperon}},
  \bibinfo {author} {\bibfnamefont {G.}~\bibnamefont {Grancini}}, \bibinfo
  {author} {\bibfnamefont {C.}~\bibnamefont {Menelaou}}, \bibinfo {author}
  {\bibfnamefont {M.~J.~P.}\ \bibnamefont {Alcocer}}, \bibinfo {author}
  {\bibfnamefont {T.}~\bibnamefont {Leijtens}}, \bibinfo {author}
  {\bibfnamefont {L.~M.}\ \bibnamefont {Herz}}, \bibinfo {author}
  {\bibfnamefont {A.}~\bibnamefont {Petrozza}},\ and\ \bibinfo {author}
  {\bibfnamefont {H.~J.}\ \bibnamefont {Snaith}},\ }\href
  {https://doi.org/10.1126/science.1243982} {\bibfield  {journal} {\bibinfo
  {journal} {Science}\ }\textbf {\bibinfo {volume} {342}},\ \bibinfo {pages}
  {341} (\bibinfo {year} {2013})}\BibitemShut {NoStop}%
\bibitem [{\citenamefont {Shi}\ \emph {et~al.}(2015)\citenamefont {Shi},
  \citenamefont {Adinolfi}, \citenamefont {Comin}, \citenamefont {Yuan},
  \citenamefont {Alarousu}, \citenamefont {Buin}, \citenamefont {Chen},
  \citenamefont {Hoogland}, \citenamefont {Rothenberger}, \citenamefont
  {Katsiev}, \citenamefont {Losovyj}, \citenamefont {Zhang}, \citenamefont
  {Dowben}, \citenamefont {Mohammed}, \citenamefont {Sargent},\ and\
  \citenamefont {Bakr}}]{Shi2015}%
  \BibitemOpen
  \bibfield  {author} {\bibinfo {author} {\bibfnamefont {D.}~\bibnamefont
  {Shi}}, \bibinfo {author} {\bibfnamefont {V.}~\bibnamefont {Adinolfi}},
  \bibinfo {author} {\bibfnamefont {R.}~\bibnamefont {Comin}}, \bibinfo
  {author} {\bibfnamefont {M.}~\bibnamefont {Yuan}}, \bibinfo {author}
  {\bibfnamefont {E.}~\bibnamefont {Alarousu}}, \bibinfo {author}
  {\bibfnamefont {A.}~\bibnamefont {Buin}}, \bibinfo {author} {\bibfnamefont
  {Y.}~\bibnamefont {Chen}}, \bibinfo {author} {\bibfnamefont {S.}~\bibnamefont
  {Hoogland}}, \bibinfo {author} {\bibfnamefont {A.}~\bibnamefont
  {Rothenberger}}, \bibinfo {author} {\bibfnamefont {K.}~\bibnamefont
  {Katsiev}}, \bibinfo {author} {\bibfnamefont {Y.}~\bibnamefont {Losovyj}},
  \bibinfo {author} {\bibfnamefont {X.}~\bibnamefont {Zhang}}, \bibinfo
  {author} {\bibfnamefont {P.~A.}\ \bibnamefont {Dowben}}, \bibinfo {author}
  {\bibfnamefont {O.~F.}\ \bibnamefont {Mohammed}}, \bibinfo {author}
  {\bibfnamefont {E.~H.}\ \bibnamefont {Sargent}},\ and\ \bibinfo {author}
  {\bibfnamefont {O.~M.}\ \bibnamefont {Bakr}},\ }\href
  {https://doi.org/10.1126/science.aaa2725} {\bibfield  {journal} {\bibinfo
  {journal} {Science}\ }\textbf {\bibinfo {volume} {347}},\ \bibinfo {pages}
  {519} (\bibinfo {year} {2015})}\BibitemShut {NoStop}%
\bibitem [{\citenamefont {Chen}\ \emph {et~al.}(2016)\citenamefont {Chen},
  \citenamefont {Yi}, \citenamefont {Wu}, \citenamefont {Haroldson},
  \citenamefont {Gartstein}, \citenamefont {Rodionov}, \citenamefont
  {Tikhonov}, \citenamefont {Zakhidov}, \citenamefont {Zhu},\ and\
  \citenamefont {Podzorov}}]{Chen2016}%
  \BibitemOpen
  \bibfield  {author} {\bibinfo {author} {\bibfnamefont {Y.}~\bibnamefont
  {Chen}}, \bibinfo {author} {\bibfnamefont {H.~T.}\ \bibnamefont {Yi}},
  \bibinfo {author} {\bibfnamefont {X.}~\bibnamefont {Wu}}, \bibinfo {author}
  {\bibfnamefont {R.}~\bibnamefont {Haroldson}}, \bibinfo {author}
  {\bibfnamefont {Y.~N.}\ \bibnamefont {Gartstein}}, \bibinfo {author}
  {\bibfnamefont {Y.~I.}\ \bibnamefont {Rodionov}}, \bibinfo {author}
  {\bibfnamefont {K.~S.}\ \bibnamefont {Tikhonov}}, \bibinfo {author}
  {\bibfnamefont {A.}~\bibnamefont {Zakhidov}}, \bibinfo {author}
  {\bibfnamefont {X.~Y.}\ \bibnamefont {Zhu}},\ and\ \bibinfo {author}
  {\bibfnamefont {V.}~\bibnamefont {Podzorov}},\ }\href
  {https://doi.org/10.1038/ncomms12253} {\bibfield  {journal} {\bibinfo
  {journal} {Nat. Commun.}\ }\textbf {\bibinfo {volume} {7}},\ \bibinfo {pages}
  {12253} (\bibinfo {year} {2016})}\BibitemShut {NoStop}%
\bibitem [{\citenamefont {Herz}(2017)}]{Herz2017}%
  \BibitemOpen
  \bibfield  {author} {\bibinfo {author} {\bibfnamefont {L.~M.}\ \bibnamefont
  {Herz}},\ }\href {https://doi.org/10.1021/acsenergylett.7b00276} {\bibfield
  {journal} {\bibinfo  {journal} {ACS Energy Lett.}\ }\textbf {\bibinfo
  {volume} {2}},\ \bibinfo {pages} {1539} (\bibinfo {year} {2017})}\BibitemShut
  {NoStop}%
\bibitem [{\citenamefont {Jena}\ \emph {et~al.}(2019)\citenamefont {Jena},
  \citenamefont {Kulkarni},\ and\ \citenamefont {Miyasaka}}]{Jena2019}%
  \BibitemOpen
  \bibfield  {author} {\bibinfo {author} {\bibfnamefont {A.~K.}\ \bibnamefont
  {Jena}}, \bibinfo {author} {\bibfnamefont {A.}~\bibnamefont {Kulkarni}},\
  and\ \bibinfo {author} {\bibfnamefont {T.}~\bibnamefont {Miyasaka}},\ }\href
  {https://doi.org/10.1021/acs.chemrev.8b00539} {\bibfield  {journal} {\bibinfo
   {journal} {Chem. Rev.}\ }\textbf {\bibinfo {volume} {119}},\ \bibinfo
  {pages} {3036} (\bibinfo {year} {2019})}\BibitemShut {NoStop}%
\bibitem [{\citenamefont {Beecher}\ \emph {et~al.}(2016)\citenamefont
  {Beecher}, \citenamefont {Semonin}, \citenamefont {Skelton}, \citenamefont
  {Frost}, \citenamefont {Terban}, \citenamefont {Zhai}, \citenamefont
  {Alatas}, \citenamefont {Owen}, \citenamefont {Walsh},\ and\ \citenamefont
  {Billinge}}]{Beecher2016}%
  \BibitemOpen
  \bibfield  {author} {\bibinfo {author} {\bibfnamefont {A.~N.}\ \bibnamefont
  {Beecher}}, \bibinfo {author} {\bibfnamefont {O.~E.}\ \bibnamefont
  {Semonin}}, \bibinfo {author} {\bibfnamefont {J.~M.}\ \bibnamefont
  {Skelton}}, \bibinfo {author} {\bibfnamefont {J.~M.}\ \bibnamefont {Frost}},
  \bibinfo {author} {\bibfnamefont {M.~W.}\ \bibnamefont {Terban}}, \bibinfo
  {author} {\bibfnamefont {H.}~\bibnamefont {Zhai}}, \bibinfo {author}
  {\bibfnamefont {A.}~\bibnamefont {Alatas}}, \bibinfo {author} {\bibfnamefont
  {J.~S.}\ \bibnamefont {Owen}}, \bibinfo {author} {\bibfnamefont
  {A.}~\bibnamefont {Walsh}},\ and\ \bibinfo {author} {\bibfnamefont {S.~J.}\
  \bibnamefont {Billinge}},\ }\href
  {https://doi.org/10.1021/acsenergylett.6b00381} {\bibfield  {journal}
  {\bibinfo  {journal} {ACS Energy Lett.}\ }\textbf {\bibinfo {volume} {1}},\
  \bibinfo {pages} {880} (\bibinfo {year} {2016})}\BibitemShut {NoStop}%
\bibitem [{\citenamefont {Yaffe}\ \emph {et~al.}(2017)\citenamefont {Yaffe},
  \citenamefont {Guo}, \citenamefont {Tan}, \citenamefont {Egger},
  \citenamefont {Hull}, \citenamefont {Stoumpos}, \citenamefont {Zheng},
  \citenamefont {Heinz}, \citenamefont {Kronik}, \citenamefont {Kanatzidis},
  \citenamefont {Owen}, \citenamefont {Rappe}, \citenamefont {Pimenta},\ and\
  \citenamefont {Brus}}]{Yaffe2017}%
  \BibitemOpen
  \bibfield  {author} {\bibinfo {author} {\bibfnamefont {O.}~\bibnamefont
  {Yaffe}}, \bibinfo {author} {\bibfnamefont {Y.}~\bibnamefont {Guo}}, \bibinfo
  {author} {\bibfnamefont {L.~Z.}\ \bibnamefont {Tan}}, \bibinfo {author}
  {\bibfnamefont {D.~A.}\ \bibnamefont {Egger}}, \bibinfo {author}
  {\bibfnamefont {T.}~\bibnamefont {Hull}}, \bibinfo {author} {\bibfnamefont
  {C.~C.}\ \bibnamefont {Stoumpos}}, \bibinfo {author} {\bibfnamefont
  {F.}~\bibnamefont {Zheng}}, \bibinfo {author} {\bibfnamefont {T.~F.}\
  \bibnamefont {Heinz}}, \bibinfo {author} {\bibfnamefont {L.}~\bibnamefont
  {Kronik}}, \bibinfo {author} {\bibfnamefont {M.~G.}\ \bibnamefont
  {Kanatzidis}}, \bibinfo {author} {\bibfnamefont {J.~S.}\ \bibnamefont
  {Owen}}, \bibinfo {author} {\bibfnamefont {A.~M.}\ \bibnamefont {Rappe}},
  \bibinfo {author} {\bibfnamefont {M.~A.}\ \bibnamefont {Pimenta}},\ and\
  \bibinfo {author} {\bibfnamefont {L.~E.}\ \bibnamefont {Brus}},\ }\href
  {https://doi.org/10.1103/PhysRevLett.118.136001} {\bibfield  {journal}
  {\bibinfo  {journal} {Phys. Rev. Lett.}\ }\textbf {\bibinfo {volume} {118}},\
  \bibinfo {pages} {136001} (\bibinfo {year} {2017})}\BibitemShut {NoStop}%
\bibitem [{\citenamefont {Miyata}\ and\ \citenamefont
  {Zhu}(2018)}]{Miyata2018}%
  \BibitemOpen
  \bibfield  {author} {\bibinfo {author} {\bibfnamefont {K.}~\bibnamefont
  {Miyata}}\ and\ \bibinfo {author} {\bibfnamefont {X.-Y.}\ \bibnamefont
  {Zhu}},\ }\href {https://doi.org/10.1038/s41563-018-0068-7} {\bibfield
  {journal} {\bibinfo  {journal} {Nat. Mater.}\ }\textbf {\bibinfo {volume}
  {17}},\ \bibinfo {pages} {379} (\bibinfo {year} {2018})}\BibitemShut
  {NoStop}%
\bibitem [{\citenamefont {Katan}\ \emph {et~al.}(2018)\citenamefont {Katan},
  \citenamefont {Mohite},\ and\ \citenamefont {Even}}]{Katan2018}%
  \BibitemOpen
  \bibfield  {author} {\bibinfo {author} {\bibfnamefont {C.}~\bibnamefont
  {Katan}}, \bibinfo {author} {\bibfnamefont {A.~D.}\ \bibnamefont {Mohite}},\
  and\ \bibinfo {author} {\bibfnamefont {J.}~\bibnamefont {Even}},\ }\href
  {https://doi.org/10.1038/s41563-018-0070-0} {\bibfield  {journal} {\bibinfo
  {journal} {Nat. Mater.}\ }\textbf {\bibinfo {volume} {17}},\ \bibinfo {pages}
  {377} (\bibinfo {year} {2018})}\BibitemShut {NoStop}%
\bibitem [{\citenamefont {Mayers}\ \emph {et~al.}(2018)\citenamefont {Mayers},
  \citenamefont {Tan}, \citenamefont {Egger}, \citenamefont {Rappe},\ and\
  \citenamefont {Reichman}}]{Mayers2018}%
  \BibitemOpen
  \bibfield  {author} {\bibinfo {author} {\bibfnamefont {M.~Z.}\ \bibnamefont
  {Mayers}}, \bibinfo {author} {\bibfnamefont {L.~Z.}\ \bibnamefont {Tan}},
  \bibinfo {author} {\bibfnamefont {D.~A.}\ \bibnamefont {Egger}}, \bibinfo
  {author} {\bibfnamefont {A.~M.}\ \bibnamefont {Rappe}},\ and\ \bibinfo
  {author} {\bibfnamefont {D.~R.}\ \bibnamefont {Reichman}},\ }\href
  {https://doi.org/10.1021/acs.nanolett.8b04276} {\bibfield  {journal}
  {\bibinfo  {journal} {Nano Lett.}\ }\textbf {\bibinfo {volume} {18}},\
  \bibinfo {pages} {8041} (\bibinfo {year} {2018})}\BibitemShut {NoStop}%
\bibitem [{\citenamefont {Li}\ \emph {et~al.}(2019)\citenamefont {Li},
  \citenamefont {Vasenko}, \citenamefont {Tang},\ and\ \citenamefont
  {Prezhdo}}]{Li2019h}%
  \BibitemOpen
  \bibfield  {author} {\bibinfo {author} {\bibfnamefont {W.}~\bibnamefont
  {Li}}, \bibinfo {author} {\bibfnamefont {A.~S.}\ \bibnamefont {Vasenko}},
  \bibinfo {author} {\bibfnamefont {J.}~\bibnamefont {Tang}},\ and\ \bibinfo
  {author} {\bibfnamefont {O.~V.}\ \bibnamefont {Prezhdo}},\ }\href
  {https://doi.org/10.1021/acs.jpclett.9b02553} {\bibfield  {journal} {\bibinfo
   {journal} {J. Phys. Chem. Lett.}\ }\textbf {\bibinfo {volume} {10}},\
  \bibinfo {pages} {6219} (\bibinfo {year} {2019})}\BibitemShut {NoStop}%
\bibitem [{\citenamefont {Ferreira}\ \emph {et~al.}(2020)\citenamefont
  {Ferreira}, \citenamefont {Paofai}, \citenamefont {L{\'{e}}toublon},
  \citenamefont {Ollivier}, \citenamefont {Raymond}, \citenamefont {Hehlen},
  \citenamefont {Ruffl{\'{e}}}, \citenamefont {Cordier}, \citenamefont {Katan},
  \citenamefont {Even},\ and\ \citenamefont {Bourges}}]{Ferreira2020}%
  \BibitemOpen
  \bibfield  {author} {\bibinfo {author} {\bibfnamefont {A.~C.}\ \bibnamefont
  {Ferreira}}, \bibinfo {author} {\bibfnamefont {S.}~\bibnamefont {Paofai}},
  \bibinfo {author} {\bibfnamefont {A.}~\bibnamefont {L{\'{e}}toublon}},
  \bibinfo {author} {\bibfnamefont {J.}~\bibnamefont {Ollivier}}, \bibinfo
  {author} {\bibfnamefont {S.}~\bibnamefont {Raymond}}, \bibinfo {author}
  {\bibfnamefont {B.}~\bibnamefont {Hehlen}}, \bibinfo {author} {\bibfnamefont
  {B.}~\bibnamefont {Ruffl{\'{e}}}}, \bibinfo {author} {\bibfnamefont
  {S.}~\bibnamefont {Cordier}}, \bibinfo {author} {\bibfnamefont
  {C.}~\bibnamefont {Katan}}, \bibinfo {author} {\bibfnamefont
  {J.}~\bibnamefont {Even}},\ and\ \bibinfo {author} {\bibfnamefont
  {P.}~\bibnamefont {Bourges}},\ }\href
  {https://doi.org/10.1038/s42005-020-0313-7} {\bibfield  {journal} {\bibinfo
  {journal} {Commun. Phys.}\ }\textbf {\bibinfo {volume} {3}},\ \bibinfo
  {pages} {48} (\bibinfo {year} {2020})}\BibitemShut {NoStop}%
\bibitem [{\citenamefont {Wang}\ \emph {et~al.}(2021)\citenamefont {Wang},
  \citenamefont {Fu}, \citenamefont {Ziffer}, \citenamefont {Dai},
  \citenamefont {Maehrlein},\ and\ \citenamefont {Zhu}}]{Wang2021e}%
  \BibitemOpen
  \bibfield  {author} {\bibinfo {author} {\bibfnamefont {F.}~\bibnamefont
  {Wang}}, \bibinfo {author} {\bibfnamefont {Y.}~\bibnamefont {Fu}}, \bibinfo
  {author} {\bibfnamefont {M.~E.}\ \bibnamefont {Ziffer}}, \bibinfo {author}
  {\bibfnamefont {Y.}~\bibnamefont {Dai}}, \bibinfo {author} {\bibfnamefont
  {S.~F.}\ \bibnamefont {Maehrlein}},\ and\ \bibinfo {author} {\bibfnamefont
  {X.-Y.}\ \bibnamefont {Zhu}},\ }\href {https://doi.org/10.1021/jacs.0c10943}
  {\bibfield  {journal} {\bibinfo  {journal} {J. Am. Chem. Soc.}\ }\textbf
  {\bibinfo {volume} {143}},\ \bibinfo {pages} {5} (\bibinfo {year}
  {2021})}\BibitemShut {NoStop}%
\bibitem [{\citenamefont {Lanigan-Atkins}\ \emph {et~al.}(2021)\citenamefont
  {Lanigan-Atkins}, \citenamefont {He}, \citenamefont {Krogstad}, \citenamefont
  {Pajerowski}, \citenamefont {Abernathy}, \citenamefont {Xu}, \citenamefont
  {Xu}, \citenamefont {Chung}, \citenamefont {Kanatzidis}, \citenamefont
  {Rosenkranz}, \citenamefont {Osborn},\ and\ \citenamefont
  {Delaire}}]{Lanigan-Atkins2021}%
  \BibitemOpen
  \bibfield  {author} {\bibinfo {author} {\bibfnamefont {T.}~\bibnamefont
  {Lanigan-Atkins}}, \bibinfo {author} {\bibfnamefont {X.}~\bibnamefont {He}},
  \bibinfo {author} {\bibfnamefont {M.~J.}\ \bibnamefont {Krogstad}}, \bibinfo
  {author} {\bibfnamefont {D.~M.}\ \bibnamefont {Pajerowski}}, \bibinfo
  {author} {\bibfnamefont {D.~L.}\ \bibnamefont {Abernathy}}, \bibinfo {author}
  {\bibfnamefont {G.~N. M.~N.}\ \bibnamefont {Xu}}, \bibinfo {author}
  {\bibfnamefont {Z.}~\bibnamefont {Xu}}, \bibinfo {author} {\bibfnamefont
  {D.-Y.}\ \bibnamefont {Chung}}, \bibinfo {author} {\bibfnamefont {M.~G.}\
  \bibnamefont {Kanatzidis}}, \bibinfo {author} {\bibfnamefont
  {S.}~\bibnamefont {Rosenkranz}}, \bibinfo {author} {\bibfnamefont
  {R.}~\bibnamefont {Osborn}},\ and\ \bibinfo {author} {\bibfnamefont
  {O.}~\bibnamefont {Delaire}},\ }\href
  {https://doi.org/10.1038/s41563-021-00947-y} {\bibfield  {journal} {\bibinfo
  {journal} {Nat. Mater.}\ }\textbf {\bibinfo {volume} {20}},\ \bibinfo {pages}
  {977} (\bibinfo {year} {2021})}\BibitemShut {NoStop}%
\bibitem [{\citenamefont {Tadano}\ and\ \citenamefont
  {Saidi}(2022)}]{Tadano2022}%
  \BibitemOpen
  \bibfield  {author} {\bibinfo {author} {\bibfnamefont {T.}~\bibnamefont
  {Tadano}}\ and\ \bibinfo {author} {\bibfnamefont {W.~A.}\ \bibnamefont
  {Saidi}},\ }\href {https://doi.org/10.1103/PhysRevLett.129.185901} {\bibfield
   {journal} {\bibinfo  {journal} {Phys. Rev. Lett.}\ }\textbf {\bibinfo
  {volume} {129}},\ \bibinfo {pages} {185901} (\bibinfo {year}
  {2022})}\BibitemShut {NoStop}%
\bibitem [{\citenamefont {Schilcher}\ \emph {et~al.}(2023)\citenamefont
  {Schilcher}, \citenamefont {Abramovitch}, \citenamefont {Mayers},
  \citenamefont {Tan}, \citenamefont {Reichman},\ and\ \citenamefont
  {Egger}}]{Schilcher2023a}%
  \BibitemOpen
  \bibfield  {author} {\bibinfo {author} {\bibfnamefont {M.~J.}\ \bibnamefont
  {Schilcher}}, \bibinfo {author} {\bibfnamefont {D.~J.}\ \bibnamefont
  {Abramovitch}}, \bibinfo {author} {\bibfnamefont {M.~Z.}\ \bibnamefont
  {Mayers}}, \bibinfo {author} {\bibfnamefont {L.~Z.}\ \bibnamefont {Tan}},
  \bibinfo {author} {\bibfnamefont {D.~R.}\ \bibnamefont {Reichman}},\ and\
  \bibinfo {author} {\bibfnamefont {D.~A.}\ \bibnamefont {Egger}},\ }\href
  {https://doi.org/10.1103/PhysRevMaterials.7.L081601} {\bibfield  {journal}
  {\bibinfo  {journal} {Phys. Rev. Mater.}\ }\textbf {\bibinfo {volume} {7}},\
  \bibinfo {pages} {L081601} (\bibinfo {year} {2023})}\BibitemShut {NoStop}%
\bibitem [{\citenamefont {Fransson}\ \emph {et~al.}(2023)\citenamefont
  {Fransson}, \citenamefont {Rosander}, \citenamefont {Eriksson}, \citenamefont
  {Rahm}, \citenamefont {Tadano},\ and\ \citenamefont {Erhart}}]{Fransson2023}%
  \BibitemOpen
  \bibfield  {author} {\bibinfo {author} {\bibfnamefont {E.}~\bibnamefont
  {Fransson}}, \bibinfo {author} {\bibfnamefont {P.}~\bibnamefont {Rosander}},
  \bibinfo {author} {\bibfnamefont {F.}~\bibnamefont {Eriksson}}, \bibinfo
  {author} {\bibfnamefont {J.~M.}\ \bibnamefont {Rahm}}, \bibinfo {author}
  {\bibfnamefont {T.}~\bibnamefont {Tadano}},\ and\ \bibinfo {author}
  {\bibfnamefont {P.}~\bibnamefont {Erhart}},\ }\href
  {https://doi.org/10.1038/s42005-023-01297-8} {\bibfield  {journal} {\bibinfo
  {journal} {Commun. Phys.}\ }\textbf {\bibinfo {volume} {6}},\ \bibinfo
  {pages} {173} (\bibinfo {year} {2023})}\BibitemShut {NoStop}%
\bibitem [{\citenamefont {Kubo}(1969)}]{Kubo1969}%
  \BibitemOpen
  \bibfield  {author} {\bibinfo {author} {\bibfnamefont {R.}~\bibnamefont
  {Kubo}},\ }\href
  {https://onlinelibrary.wiley.com/doi/abs/10.1002/9780470143605.ch6}
  {\bibfield  {journal} {\bibinfo  {journal} {Adv. Chem. Phys.}\ }\textbf
  {\bibinfo {volume} {15}},\ \bibinfo {pages} {101} (\bibinfo {year}
  {1969})}\BibitemShut {NoStop}%
\bibitem [{\citenamefont {Migchelsen}\ and\ \citenamefont
  {Berendsen}(1973)}]{Migchelsen1973}%
  \BibitemOpen
  \bibfield  {author} {\bibinfo {author} {\bibfnamefont {C.}~\bibnamefont
  {Migchelsen}}\ and\ \bibinfo {author} {\bibfnamefont {H.~J.~C.}\ \bibnamefont
  {Berendsen}},\ }\href {https://doi.org/10.1063/1.1679805} {\bibfield
  {journal} {\bibinfo  {journal} {J. Chem. Phys.}\ }\textbf {\bibinfo {volume}
  {59}},\ \bibinfo {pages} {296} (\bibinfo {year} {1973})}\BibitemShut
  {NoStop}%
\bibitem [{\citenamefont {Jeener}\ \emph {et~al.}(1979)\citenamefont {Jeener},
  \citenamefont {Meier}, \citenamefont {Bachmann},\ and\ \citenamefont
  {Ernst}}]{Jeener1979}%
  \BibitemOpen
  \bibfield  {author} {\bibinfo {author} {\bibfnamefont {J.}~\bibnamefont
  {Jeener}}, \bibinfo {author} {\bibfnamefont {B.~H.}\ \bibnamefont {Meier}},
  \bibinfo {author} {\bibfnamefont {P.}~\bibnamefont {Bachmann}},\ and\
  \bibinfo {author} {\bibfnamefont {R.~R.}\ \bibnamefont {Ernst}},\ }\href
  {https://doi.org/10.1063/1.438208} {\bibfield  {journal} {\bibinfo  {journal}
  {J. Chem. Phys.}\ }\textbf {\bibinfo {volume} {71}},\ \bibinfo {pages} {4546}
  (\bibinfo {year} {1979})}\BibitemShut {NoStop}%
\bibitem [{\citenamefont {{\v{S}}anda}\ and\ \citenamefont
  {Mukamel}(2006)}]{Sanda2006}%
  \BibitemOpen
  \bibfield  {author} {\bibinfo {author} {\bibfnamefont {F.}~\bibnamefont
  {{\v{S}}anda}}\ and\ \bibinfo {author} {\bibfnamefont {S.}~\bibnamefont
  {Mukamel}},\ }\href {https://doi.org/10.1103/PhysRevE.73.011103} {\bibfield
  {journal} {\bibinfo  {journal} {Phys. Rev. E}\ }\textbf {\bibinfo {volume}
  {73}},\ \bibinfo {pages} {011103} (\bibinfo {year} {2006})}\BibitemShut
  {NoStop}%
\bibitem [{\citenamefont {{L. C. Jansen}}\ and\ \citenamefont
  {Knoester}(2007)}]{Jansen2007}%
  \BibitemOpen
  \bibfield  {author} {\bibinfo {author} {\bibfnamefont {T.}~\bibnamefont {{L.
  C. Jansen}}}\ and\ \bibinfo {author} {\bibfnamefont {J.}~\bibnamefont
  {Knoester}},\ }\href {https://doi.org/10.1063/1.2806179} {\bibfield
  {journal} {\bibinfo  {journal} {J. Chem. Phys.}\ }\textbf {\bibinfo {volume}
  {127}},\ \bibinfo {pages} {234502} (\bibinfo {year} {2007})}\BibitemShut
  {NoStop}%
\bibitem [{\citenamefont {Tamura}\ \emph {et~al.}(2008)\citenamefont {Tamura},
  \citenamefont {Ramon}, \citenamefont {Bittner},\ and\ \citenamefont
  {Burghardt}}]{Tamura2008}%
  \BibitemOpen
  \bibfield  {author} {\bibinfo {author} {\bibfnamefont {H.}~\bibnamefont
  {Tamura}}, \bibinfo {author} {\bibfnamefont {J.~G.~S.}\ \bibnamefont
  {Ramon}}, \bibinfo {author} {\bibfnamefont {E.~R.}\ \bibnamefont {Bittner}},\
  and\ \bibinfo {author} {\bibfnamefont {I.}~\bibnamefont {Burghardt}},\ }\href
  {https://doi.org/10.1103/PhysRevLett.100.107402} {\bibfield  {journal}
  {\bibinfo  {journal} {Phys. Rev. Lett.}\ }\textbf {\bibinfo {volume} {100}},\
  \bibinfo {pages} {107402} (\bibinfo {year} {2008})}\BibitemShut {NoStop}%
\bibitem [{\citenamefont {Tamura}\ and\ \citenamefont
  {Burghardt}(2013)}]{Tamura2013}%
  \BibitemOpen
  \bibfield  {author} {\bibinfo {author} {\bibfnamefont {H.}~\bibnamefont
  {Tamura}}\ and\ \bibinfo {author} {\bibfnamefont {I.}~\bibnamefont
  {Burghardt}},\ }\href {https://doi.org/10.1021/ja4093874} {\bibfield
  {journal} {\bibinfo  {journal} {J. Am. Chem. Soc.}\ }\textbf {\bibinfo
  {volume} {135}},\ \bibinfo {pages} {16364} (\bibinfo {year}
  {2013})}\BibitemShut {NoStop}%
\bibitem [{\citenamefont {Smith}\ and\ \citenamefont {Chin}(2014)}]{Smith2014}%
  \BibitemOpen
  \bibfield  {author} {\bibinfo {author} {\bibfnamefont {S.~L.}\ \bibnamefont
  {Smith}}\ and\ \bibinfo {author} {\bibfnamefont {A.~W.}\ \bibnamefont
  {Chin}},\ }\href {https://doi.org/10.1039/C4CP01791A} {\bibfield  {journal}
  {\bibinfo  {journal} {Phys. Chem. Chem. Phys.}\ }\textbf {\bibinfo {volume}
  {16}},\ \bibinfo {pages} {20305} (\bibinfo {year} {2014})}\BibitemShut
  {NoStop}%
\bibitem [{\citenamefont {Smith}\ and\ \citenamefont {Chin}(2015)}]{Smith2015}%
  \BibitemOpen
  \bibfield  {author} {\bibinfo {author} {\bibfnamefont {S.~L.}\ \bibnamefont
  {Smith}}\ and\ \bibinfo {author} {\bibfnamefont {A.~W.}\ \bibnamefont
  {Chin}},\ }\href {https://doi.org/10.1103/PhysRevB.91.201302} {\bibfield
  {journal} {\bibinfo  {journal} {Phys. Rev. B}\ }\textbf {\bibinfo {volume}
  {91}},\ \bibinfo {pages} {201302} (\bibinfo {year} {2015})}\BibitemShut
  {NoStop}%
\bibitem [{\citenamefont {Kato}\ and\ \citenamefont
  {Ishizaki}(2018)}]{Kato2018b}%
  \BibitemOpen
  \bibfield  {author} {\bibinfo {author} {\bibfnamefont {A.}~\bibnamefont
  {Kato}}\ and\ \bibinfo {author} {\bibfnamefont {A.}~\bibnamefont
  {Ishizaki}},\ }\href {https://doi.org/10.1103/PhysRevLett.121.026001}
  {\bibfield  {journal} {\bibinfo  {journal} {Phys. Rev. Lett.}\ }\textbf
  {\bibinfo {volume} {121}},\ \bibinfo {pages} {026001} (\bibinfo {year}
  {2018})}\BibitemShut {NoStop}%
\bibitem [{\citenamefont {Yan}\ \emph {et~al.}(2018)\citenamefont {Yan},
  \citenamefont {Song},\ and\ \citenamefont {Shi}}]{Yan2018}%
  \BibitemOpen
  \bibfield  {author} {\bibinfo {author} {\bibfnamefont {Y.}~\bibnamefont
  {Yan}}, \bibinfo {author} {\bibfnamefont {L.}~\bibnamefont {Song}},\ and\
  \bibinfo {author} {\bibfnamefont {Q.}~\bibnamefont {Shi}},\ }\href
  {https://doi.org/10.1063/1.5017866} {\bibfield  {journal} {\bibinfo
  {journal} {J. Chem. Phys.}\ }\textbf {\bibinfo {volume} {148}},\ \bibinfo
  {pages} {084109} (\bibinfo {year} {2018})}\BibitemShut {NoStop}%
\bibitem [{\citenamefont {Yan}\ \emph {et~al.}(2019)\citenamefont {Yan},
  \citenamefont {Xu}, \citenamefont {Liu},\ and\ \citenamefont
  {Shi}}]{Yan2019}%
  \BibitemOpen
  \bibfield  {author} {\bibinfo {author} {\bibfnamefont {Y.}~\bibnamefont
  {Yan}}, \bibinfo {author} {\bibfnamefont {M.}~\bibnamefont {Xu}}, \bibinfo
  {author} {\bibfnamefont {Y.}~\bibnamefont {Liu}},\ and\ \bibinfo {author}
  {\bibfnamefont {Q.}~\bibnamefont {Shi}},\ }\href
  {https://doi.org/10.1063/1.5096214} {\bibfield  {journal} {\bibinfo
  {journal} {J. Chem. Phys.}\ }\textbf {\bibinfo {volume} {150}},\ \bibinfo
  {pages} {234101} (\bibinfo {year} {2019})}\BibitemShut {NoStop}%
\bibitem [{\citenamefont {Fetherolf}\ \emph {et~al.}(2020)\citenamefont
  {Fetherolf}, \citenamefont {Gole{\v{z}}},\ and\ \citenamefont
  {Berkelbach}}]{Fetherolf2020}%
  \BibitemOpen
  \bibfield  {author} {\bibinfo {author} {\bibfnamefont {J.~H.}\ \bibnamefont
  {Fetherolf}}, \bibinfo {author} {\bibfnamefont {D.}~\bibnamefont
  {Gole{\v{z}}}},\ and\ \bibinfo {author} {\bibfnamefont {T.~C.}\ \bibnamefont
  {Berkelbach}},\ }\href {https://doi.org/10.1103/PhysRevX.10.021062}
  {\bibfield  {journal} {\bibinfo  {journal} {Phys. Rev. X}\ }\textbf {\bibinfo
  {volume} {10}},\ \bibinfo {pages} {021062} (\bibinfo {year}
  {2020})}\BibitemShut {NoStop}%
\bibitem [{\citenamefont {Balzer}\ and\ \citenamefont
  {Kassal}(2022)}]{Balzer2022}%
  \BibitemOpen
  \bibfield  {author} {\bibinfo {author} {\bibfnamefont {D.}~\bibnamefont
  {Balzer}}\ and\ \bibinfo {author} {\bibfnamefont {I.}~\bibnamefont
  {Kassal}},\ }\href {https://doi.org/10.1126/sciadv.abl9692} {\bibfield
  {journal} {\bibinfo  {journal} {Sci. Adv.}\ }\textbf {\bibinfo {volume}
  {8}},\ \bibinfo {pages} {eabl9692} (\bibinfo {year} {2022})}\BibitemShut
  {NoStop}%
\bibitem [{\citenamefont {Fetherolf}\ \emph {et~al.}(2023)\citenamefont
  {Fetherolf}, \citenamefont {Shih},\ and\ \citenamefont
  {Berkelbach}}]{Fetherolf2023}%
  \BibitemOpen
  \bibfield  {author} {\bibinfo {author} {\bibfnamefont {J.~H.}\ \bibnamefont
  {Fetherolf}}, \bibinfo {author} {\bibfnamefont {P.}~\bibnamefont {Shih}},\
  and\ \bibinfo {author} {\bibfnamefont {T.~C.}\ \bibnamefont {Berkelbach}},\
  }\href {https://doi.org/10.1103/PhysRevB.107.064304} {\bibfield  {journal}
  {\bibinfo  {journal} {Phys. Rev. B}\ }\textbf {\bibinfo {volume} {107}},\
  \bibinfo {pages} {064304} (\bibinfo {year} {2023})}\BibitemShut {NoStop}%
\bibitem [{\citenamefont {Nakajima}(1958)}]{Nakajima1958}%
  \BibitemOpen
  \bibfield  {author} {\bibinfo {author} {\bibfnamefont {S.}~\bibnamefont
  {Nakajima}},\ }\href {https://doi.org/10.1143/PTP.20.948} {\bibfield
  {journal} {\bibinfo  {journal} {Prog. Theor. Phys.}\ }\textbf {\bibinfo
  {volume} {20}},\ \bibinfo {pages} {948} (\bibinfo {year} {1958})}\BibitemShut
  {NoStop}%
\bibitem [{\citenamefont {Zwanzig}(1960)}]{Zwanzig1960}%
  \BibitemOpen
  \bibfield  {author} {\bibinfo {author} {\bibfnamefont {R.}~\bibnamefont
  {Zwanzig}},\ }\href {https://doi.org/10.1063/1.1731409} {\bibfield  {journal}
  {\bibinfo  {journal} {J. Chem. Phys.}\ }\textbf {\bibinfo {volume} {33}},\
  \bibinfo {pages} {1338} (\bibinfo {year} {1960})}\BibitemShut {NoStop}%
\bibitem [{\citenamefont {Ishizaki}\ \emph {et~al.}(2010)\citenamefont
  {Ishizaki}, \citenamefont {Calhoun}, \citenamefont {Schlau-Cohen},\ and\
  \citenamefont {Fleming}}]{Ishizaki2010}%
  \BibitemOpen
  \bibfield  {author} {\bibinfo {author} {\bibfnamefont {A.}~\bibnamefont
  {Ishizaki}}, \bibinfo {author} {\bibfnamefont {T.~R.}\ \bibnamefont
  {Calhoun}}, \bibinfo {author} {\bibfnamefont {G.~S.}\ \bibnamefont
  {Schlau-Cohen}},\ and\ \bibinfo {author} {\bibfnamefont {G.~R.}\ \bibnamefont
  {Fleming}},\ }\href {https://doi.org/10.1039/c003389h} {\bibfield  {journal}
  {\bibinfo  {journal} {Phys. Chem. Chem. Phys.}\ }\textbf {\bibinfo {volume}
  {12}},\ \bibinfo {pages} {7319} (\bibinfo {year} {2010})}\BibitemShut
  {NoStop}%
\bibitem [{\citenamefont {Yoon}\ \emph {et~al.}(1975)\citenamefont {Yoon},
  \citenamefont {Deutch},\ and\ \citenamefont {Freed}}]{Yoon1975}%
  \BibitemOpen
  \bibfield  {author} {\bibinfo {author} {\bibfnamefont {B.}~\bibnamefont
  {Yoon}}, \bibinfo {author} {\bibfnamefont {J.~M.}\ \bibnamefont {Deutch}},\
  and\ \bibinfo {author} {\bibfnamefont {J.~H.}\ \bibnamefont {Freed}},\ }\href
  {https://doi.org/10.1063/1.430417} {\bibfield  {journal} {\bibinfo  {journal}
  {J. Chem. Phys.}\ }\textbf {\bibinfo {volume} {62}},\ \bibinfo {pages} {4687}
  (\bibinfo {year} {1975})}\BibitemShut {NoStop}%
\bibitem [{\citenamefont {Yang}\ \emph {et~al.}(2017)\citenamefont {Yang},
  \citenamefont {Skelton}, \citenamefont {da~Silva}, \citenamefont {Frost},\
  and\ \citenamefont {Walsh}}]{Yang2017b}%
  \BibitemOpen
  \bibfield  {author} {\bibinfo {author} {\bibfnamefont {R.~X.}\ \bibnamefont
  {Yang}}, \bibinfo {author} {\bibfnamefont {J.~M.}\ \bibnamefont {Skelton}},
  \bibinfo {author} {\bibfnamefont {E.~L.}\ \bibnamefont {da~Silva}}, \bibinfo
  {author} {\bibfnamefont {J.~M.}\ \bibnamefont {Frost}},\ and\ \bibinfo
  {author} {\bibfnamefont {A.}~\bibnamefont {Walsh}},\ }\href
  {https://doi.org/10.1021/acs.jpclett.7b02423} {\bibfield  {journal} {\bibinfo
   {journal} {J. Phys. Chem. Lett.}\ }\textbf {\bibinfo {volume} {8}},\
  \bibinfo {pages} {4720} (\bibinfo {year} {2017})}\BibitemShut {NoStop}%
\bibitem [{\citenamefont {Marronnier}\ \emph {et~al.}(2017)\citenamefont
  {Marronnier}, \citenamefont {Lee}, \citenamefont {Geffroy}, \citenamefont
  {Even}, \citenamefont {Bonnassieux},\ and\ \citenamefont
  {Roma}}]{Marronnier2017}%
  \BibitemOpen
  \bibfield  {author} {\bibinfo {author} {\bibfnamefont {A.}~\bibnamefont
  {Marronnier}}, \bibinfo {author} {\bibfnamefont {H.}~\bibnamefont {Lee}},
  \bibinfo {author} {\bibfnamefont {B.}~\bibnamefont {Geffroy}}, \bibinfo
  {author} {\bibfnamefont {J.}~\bibnamefont {Even}}, \bibinfo {author}
  {\bibfnamefont {Y.}~\bibnamefont {Bonnassieux}},\ and\ \bibinfo {author}
  {\bibfnamefont {G.}~\bibnamefont {Roma}},\ }\href
  {https://doi.org/10.1021/acs.jpclett.7b00807} {\bibfield  {journal} {\bibinfo
   {journal} {J. Phys. Chem. Lett.}\ }\textbf {\bibinfo {volume} {8}},\
  \bibinfo {pages} {2659} (\bibinfo {year} {2017})}\BibitemShut {NoStop}%
\bibitem [{\citenamefont {Yang}\ \emph {et~al.}(2020)\citenamefont {Yang},
  \citenamefont {Skelton}, \citenamefont {da~Silva}, \citenamefont {Frost},\
  and\ \citenamefont {Walsh}}]{Yang2020c}%
  \BibitemOpen
  \bibfield  {author} {\bibinfo {author} {\bibfnamefont {R.~X.}\ \bibnamefont
  {Yang}}, \bibinfo {author} {\bibfnamefont {J.~M.}\ \bibnamefont {Skelton}},
  \bibinfo {author} {\bibfnamefont {E.~L.}\ \bibnamefont {da~Silva}}, \bibinfo
  {author} {\bibfnamefont {J.~M.}\ \bibnamefont {Frost}},\ and\ \bibinfo
  {author} {\bibfnamefont {A.}~\bibnamefont {Walsh}},\ }\href
  {https://doi.org/10.1063/1.5131575
  https://pubs.aip.org/jcp/article/152/2/024703/317569/Assessment-of-dynamic-structural-instabilities}
  {\bibfield  {journal} {\bibinfo  {journal} {J. Chem. Phys.}\ }\textbf
  {\bibinfo {volume} {152}} (\bibinfo {year} {2020})}\BibitemShut {NoStop}%
\bibitem [{\citenamefont {Wang}\ \emph {et~al.}(2022)\citenamefont {Wang},
  \citenamefont {Chu}, \citenamefont {Huber}, \citenamefont {Tu}, \citenamefont
  {Dai}, \citenamefont {Wang}, \citenamefont {Peng}, \citenamefont {Zhao},\
  and\ \citenamefont {Zhu}}]{Wang2022b}%
  \BibitemOpen
  \bibfield  {author} {\bibinfo {author} {\bibfnamefont {F.}~\bibnamefont
  {Wang}}, \bibinfo {author} {\bibfnamefont {W.}~\bibnamefont {Chu}}, \bibinfo
  {author} {\bibfnamefont {L.}~\bibnamefont {Huber}}, \bibinfo {author}
  {\bibfnamefont {T.}~\bibnamefont {Tu}}, \bibinfo {author} {\bibfnamefont
  {Y.}~\bibnamefont {Dai}}, \bibinfo {author} {\bibfnamefont {J.}~\bibnamefont
  {Wang}}, \bibinfo {author} {\bibfnamefont {H.}~\bibnamefont {Peng}}, \bibinfo
  {author} {\bibfnamefont {J.}~\bibnamefont {Zhao}},\ and\ \bibinfo {author}
  {\bibfnamefont {X.-Y.}\ \bibnamefont {Zhu}},\ }\href
  {https://doi.org/10.1073/pnas.2122436119} {\bibfield  {journal} {\bibinfo
  {journal} {Proc. Nat. Acad. Sci. USA}\ }\textbf {\bibinfo {volume} {119}},\
  \bibinfo {pages} {6} (\bibinfo {year} {2022})}\BibitemShut {NoStop}%
\bibitem [{\citenamefont {Zacharias}\ \emph {et~al.}(2023)\citenamefont
  {Zacharias}, \citenamefont {Volonakis}, \citenamefont {Giustino},\ and\
  \citenamefont {Even}}]{Zacharias2023a}%
  \BibitemOpen
  \bibfield  {author} {\bibinfo {author} {\bibfnamefont {M.}~\bibnamefont
  {Zacharias}}, \bibinfo {author} {\bibfnamefont {G.}~\bibnamefont
  {Volonakis}}, \bibinfo {author} {\bibfnamefont {F.}~\bibnamefont
  {Giustino}},\ and\ \bibinfo {author} {\bibfnamefont {J.}~\bibnamefont
  {Even}},\ }\href {https://doi.org/10.1038/s41524-023-01089-2} {\bibfield
  {journal} {\bibinfo  {journal} {npj Comput. Mater.}\ }\textbf {\bibinfo
  {volume} {9}},\ \bibinfo {pages} {153} (\bibinfo {year} {2023})}\BibitemShut
  {NoStop}%
\bibitem [{\citenamefont {Slater}(1950)}]{Slater1950}%
  \BibitemOpen
  \bibfield  {author} {\bibinfo {author} {\bibfnamefont {J.~C.}\ \bibnamefont
  {Slater}},\ }\href {https://doi.org/10.1103/PhysRev.78.748} {\bibfield
  {journal} {\bibinfo  {journal} {Phys. Rev.}\ }\textbf {\bibinfo {volume}
  {78}},\ \bibinfo {pages} {748} (\bibinfo {year} {1950})}\BibitemShut
  {NoStop}%
\bibitem [{\citenamefont {Kubo}\ \emph {et~al.}(1991)\citenamefont {Kubo},
  \citenamefont {Toda},\ and\ \citenamefont {Hashitsume}}]{Kubo1991}%
  \BibitemOpen
  \bibfield  {author} {\bibinfo {author} {\bibfnamefont {R.}~\bibnamefont
  {Kubo}}, \bibinfo {author} {\bibfnamefont {M.}~\bibnamefont {Toda}},\ and\
  \bibinfo {author} {\bibfnamefont {N.}~\bibnamefont {Hashitsume}},\ }\href
  {https://doi.org/10.1007/978-3-642-58244-8} {\emph {\bibinfo {title}
  {{Statistical Physics II: Nonequilibrium Statistical Mechanics}}}}\ (\bibinfo
   {publisher} {Springer},\ \bibinfo {address} {Berlin, Heidelberg},\ \bibinfo
  {year} {1991})\ pp.\ \bibinfo {pages} {1--279}\BibitemShut {NoStop}%
\bibitem [{\citenamefont {Weiss}(2012)}]{Weiss2012}%
  \BibitemOpen
  \bibfield  {author} {\bibinfo {author} {\bibfnamefont {U.}~\bibnamefont
  {Weiss}},\ }\href {https://doi.org/10.1142/8334} {\emph {\bibinfo {title}
  {Quantum Dissipative Systems}}},\ \bibinfo {edition} {4th}\ ed.\ (\bibinfo
  {publisher} {World Scientific},\ \bibinfo {address} {Singapore},\ \bibinfo
  {year} {2012})\ pp.\ \bibinfo {pages} {1--588}\BibitemShut {NoStop}%
\bibitem [{\citenamefont {Coleman}(2015)}]{Coleman2015}%
  \BibitemOpen
  \bibfield  {author} {\bibinfo {author} {\bibfnamefont {P.}~\bibnamefont
  {Coleman}},\ }\href {https://doi.org/10.1017/CBO9781139020916} {\emph
  {\bibinfo {title} {{Introduction to Many-Body Physics}}}}\ (\bibinfo
  {publisher} {Cambridge University Press},\ \bibinfo {year}
  {2015})\BibitemShut {NoStop}%
\bibitem [{\citenamefont {Anderson}(1954)}]{Anderson1954}%
  \BibitemOpen
  \bibfield  {author} {\bibinfo {author} {\bibfnamefont {P.~W.}\ \bibnamefont
  {Anderson}},\ }\href {https://doi.org/10.1143/JPSJ.9.316} {\bibfield
  {journal} {\bibinfo  {journal} {J. Phys. Soc. Jpn}\ }\textbf {\bibinfo
  {volume} {9}},\ \bibinfo {pages} {316} (\bibinfo {year} {1954})}\BibitemShut
  {NoStop}%
\bibitem [{\citenamefont {Kubo}(1954)}]{Kubo1954}%
  \BibitemOpen
  \bibfield  {author} {\bibinfo {author} {\bibfnamefont {R.}~\bibnamefont
  {Kubo}},\ }\href {https://doi.org/10.1143/JPSJ.9.935} {\bibfield  {journal}
  {\bibinfo  {journal} {J. Phys. Soc. Jpn.}\ }\textbf {\bibinfo {volume} {9}},\
  \bibinfo {pages} {935} (\bibinfo {year} {1954})}\BibitemShut {NoStop}%
\bibitem [{\citenamefont {Ishizaki}\ and\ \citenamefont
  {Fleming}(2009)}]{Ishizaki2009a}%
  \BibitemOpen
  \bibfield  {author} {\bibinfo {author} {\bibfnamefont {A.}~\bibnamefont
  {Ishizaki}}\ and\ \bibinfo {author} {\bibfnamefont {G.~R.}\ \bibnamefont
  {Fleming}},\ }\href {https://doi.org/10.1063/1.3155372} {\bibfield  {journal}
  {\bibinfo  {journal} {J. Chem. Phys.}\ }\textbf {\bibinfo {volume} {130}},\
  \bibinfo {pages} {234111} (\bibinfo {year} {2009})}\BibitemShut {NoStop}%
\bibitem [{\citenamefont {Prokof'ev}\ and\ \citenamefont
  {Stamp}(2000)}]{Prokofev2000}%
  \BibitemOpen
  \bibfield  {author} {\bibinfo {author} {\bibfnamefont {N.~V.}\ \bibnamefont
  {Prokof'ev}}\ and\ \bibinfo {author} {\bibfnamefont {P.~C.~E.}\ \bibnamefont
  {Stamp}},\ }\href {https://doi.org/10.1088/0034-4885/63/4/204} {\bibfield
  {journal} {\bibinfo  {journal} {Rep. Prog. Phys.}\ }\textbf {\bibinfo
  {volume} {63}},\ \bibinfo {pages} {669} (\bibinfo {year} {2000})}\BibitemShut
  {NoStop}%
\bibitem [{\citenamefont {Hsieh}\ and\ \citenamefont
  {Cao}(2018{\natexlab{a}})}]{Hsieh2018b}%
  \BibitemOpen
  \bibfield  {author} {\bibinfo {author} {\bibfnamefont {C.-Y.}\ \bibnamefont
  {Hsieh}}\ and\ \bibinfo {author} {\bibfnamefont {J.}~\bibnamefont {Cao}},\
  }\href {https://doi.org/10.1063/1.5018725} {\bibfield  {journal} {\bibinfo
  {journal} {J. Chem. Phys.}\ }\textbf {\bibinfo {volume} {148}},\ \bibinfo
  {pages} {014103} (\bibinfo {year} {2018}{\natexlab{a}})}\BibitemShut
  {NoStop}%
\bibitem [{\citenamefont {Hsieh}\ and\ \citenamefont
  {Cao}(2018{\natexlab{b}})}]{Hsieh2018c}%
  \BibitemOpen
  \bibfield  {author} {\bibinfo {author} {\bibfnamefont {C.-Y.}\ \bibnamefont
  {Hsieh}}\ and\ \bibinfo {author} {\bibfnamefont {J.}~\bibnamefont {Cao}},\
  }\href {https://doi.org/10.1063/1.5018726} {\bibfield  {journal} {\bibinfo
  {journal} {J. Chem. Phys.}\ }\textbf {\bibinfo {volume} {148}},\ \bibinfo
  {pages} {014104} (\bibinfo {year} {2018}{\natexlab{b}})}\BibitemShut
  {NoStop}%
\bibitem [{\citenamefont {Funo}\ and\ \citenamefont {Ishizaki}()}]{Funo2023}%
  \BibitemOpen
  \bibfield  {author} {\bibinfo {author} {\bibfnamefont {K.}~\bibnamefont
  {Funo}}\ and\ \bibinfo {author} {\bibfnamefont {A.}~\bibnamefont
  {Ishizaki}},\ }\href {http://arxiv.org/abs/2312.00376} {\ }\Eprint
  {https://arxiv.org/abs/2312.00376} {arXiv:2312.00376} \BibitemShut {NoStop}%
\end{thebibliography}
%

\end{document}